\documentclass[twocolumn,prb,aps]{revtex4}
\draft
\usepackage{amsmath}
\usepackage{graphics}
\newcommand{\beeq}{\begin{equation}}
\newcommand{\eneq}{\end{equation}}
\newcommand{\beeqar}{\begin{eqnarray}}
\newcommand{\eneqar}{\end{eqnarray}}
\newcommand{\op}{\boldsymbol}
\begin{document}
\title{Protective Measurements: Probing Single Quantum Systems}
\author{Tabish Qureshi}
\email{tabish@ctp-jamia.res.in}
\affiliation{Centre for Theoretical Physics, Jamia Millia Islamia, New Delhi.}
\author{N.D. Hari Dass}
\email{dass@tifrh.res.in}
\affiliation{TIFR-TCIS, Hyderabad \& CQIQC, IISc, Bangalore.}

\begin{abstract}
Making measurements on single quantum systems is considered difficult,
almost impossible if the state is a-priori unknown.
Protective measurements suggest a possibility to measure single quantum
systems and gain some new information in the process.
Protective measurement is described, both in the original and generalized
form. The degree to which the system and the apparatus remain entangled
in a protective measurement, is assessed. Possible experimental tests of
protective measurements are discussed.

\end{abstract}

\maketitle

\section{Introduction}

Quantum mechanics has been a tremendously successful theory for describing
microscopic systems. Till date there has not been a single experiment
which can demonstrate a violation of quantum theory. The success of
quantum mechanics is so widespread and robust that people have come
to believe that it is the ultimate theory for describing microscopic
systems, and that classical mechanics should be an approximation of
quantum mechanics in an appropriate limit. However, this transition from
quantum to classical has been a sticky issue since the time of the
very inception of quantum theory. 

While the state of a classical particle is adequately described by the
knowledge of its position and momentum, quantum mechanics does not
even allow simultaneous well defined values of these two quantities.
The state of a quantum particle is strangely described by a complex
entity called the wave function. Although the dynamics of state, or the
wave-function, is exactly described by the Schr\"odinger equation,
the meaning of measurable quantities takes an altogether different 
meaning. An observable described by an operator ${\mathbf A}$ is
believed to have a well defined value only if the state of the system
is an eigenstate of this operator, namely,
\begin{equation}
{\mathbf A} |a_n\rangle = a_n|a_n\rangle,
\end{equation}
where $a_n$ is called the eigenvalue of the observable. Eigenvalue
is also the outcome in a measurement of the observable ${\mathbf A}$.
Measurements of ${\mathbf A}$ on identical copies of same system,
in the state (say) $|a_k\rangle$, will all lead to the same result
$a_k$. Thus, an eigenvalue is a well defined value of ${\mathbf A}$
if the system is in its eigenstate.

However, if the system is in a state (say) $|\psi\rangle$ which is not an
eigenstate of ${\mathbf A}$, the value which one should assign to the
observable described by ${\mathbf A}$ is ambiguous. All one can do is
to define an {\em expectation value} of ${\mathbf A}$ as
\begin{equation}
\langle A\rangle  = \langle\psi|{\mathbf A}|\psi\rangle.
\end{equation}
A measurement of ${\mathbf A}$ in the state $\psi\rangle$ would still yield
an eigenvalue, one from the set $\{a_i\}$. However, the important difference
is that measurements of ${\mathbf A}$ on identical copies of same system,
in the state $|\psi\rangle$, will all lead to different eigenvalues.
The state $|\psi\rangle$ can be expanded in terms of the eigenstates of
${\mathbf A}$ as 
\begin{equation}
|\psi\rangle  = \sum_n c_n|a_n\rangle,
\end{equation}
where $c_n$ are some complex numbers. The expectation value of $A$ can now
be written as 
\begin{equation}
\langle A\rangle  = \sum_n |c_n|^2 \langle a_n|{\mathbf A}|a_n\rangle
                  = \sum_n |c_n|^2 a_n.
\end{equation}
The above expression is conventionally interpreted as average of the
measurement results on a large ensemble of identically prepared systems
in the state $|\psi\rangle$. A fraction $|c_k|^2$ of the total systems
yield the eigenvalue $a_k$.

The point to note in the above example is that while the value of $A$ is
well defined for a single system in an eigenstate $|a_k\rangle$, the
expectation value $\langle A\rangle$ in a state $|\psi\rangle$ cannot be
defined for a single system. It appears to have a meaning only for a 
large number of measurements on an ensemble of identical systems.

Issues like the one described above, still plague quantum theory, although
they are mainly interpretational issues. What sense one should make of the
formalism of quantum theory, is not clear.

\section{Quantum measurement process}

While evolution of quantum systems are very well understood, what happens
in a measurement process is not clear. This is simply because the
apparatus we use is classical, and how information from a quantum system
is carried over to the classical apparatus is not part of the quantum
formalism. Quantum theory merely postulates that in a measurement process,
the value obtained is an eigenvalue of the observable being measured, it
results in the reduction of the original state to the corresponding
eigenstate of the observable. How this process comes about is not understood,
and remains an open problem.

John von Neumann was the first one to attempt putting a quantum measurement
process on a mathematical footing.\cite{NEUMANN}
According to Neumann, a quantum measurement can be broken up into
two processes.
\begin{itemize}
\item Process 1 is a unitary process which establishes correlations
between the state of the system and state of the apparatus. It basically
correlates the various amplitudes of the system state to various
possible outcomes of the apparatus. The apparatus too has to be treated
as a quantum system. For example if the initial state $|\psi\rangle$ is
given by $|\psi\rangle = \sum_{i=1}^n c_i|a_i\rangle$ and the initial
state of the apparatus is given by $|d_0\rangle$, then the process 1
is a unitary operation
\begin{equation}
|d_0\rangle \sum_{i=1}^n c_i|a_i\rangle
 \xrightarrow[Process~1]{} \sum_{i=1}^n c_i|d_i\rangle|a_i\rangle.
\label{p1}
\end{equation}
What process 1 has done is to correlate the eigenstates of ${\mathbf A}$
with distinct states of the apparatus. The states $|d_i\rangle$ could,
for example, corresponds to some discreet positions of a pointer needle.

\item Process 2 is a non-unitary one which picks out a single outcome from
the superposition described by (\ref{p1}):
\begin{equation}
\sum_{i=1}^n c_i|d_i\rangle|a_i\rangle \xrightarrow[Process~2]{} 
|d_k\rangle|a_k\rangle.
\label{p1}
\end{equation}
It is obvious that process 2 cannot be realized through Schr\"odinger
evolution. The process 2 constitutes the heart of the so-called
{\em measurement problem}.
\end{itemize}
The mechanism behind process 2 has confounded scientists since the 
beginning of the quantum theory. It is no surprise that people have come
up with suggested resolutions which can be considered radical to fantastic
like the Everett many worlds interpretation\cite{EVERETT} or the GRW
proposal.\cite{GRW}

\subsection{Strong and (almost) impulsive measurement}

Let us first put von Neumann process 1 on a rigorous mathematical footing.
Process 1 can be constructed by a suitable interaction between the system
and the apparatus and a time evolution. Conventional quantum measurements
may be considered as the result of a strong interaction between the
system and the apparatus, active for a short duration of time. The
Hamiltonian of the system and the apparatus may be written as
\begin{equation}
\op{H}(t) = \op{H_S} + \op{H_A} + g(t)\op{Q_S} \op{Q_A} \approx g(t) \op{Q_S} \op{Q_A} ,
\end{equation}
where $H_S,~H_A$ represent the free Hamiltonians of the system and the
apparatus, respectively, and $Q_S, Q_A$ the operators of the system
and the apparatus through which they interact. We introduce another observable
$\op{R_A}$ conjugate to $\op{R_A}$, such that $[\op{R_A}, \op{Q_A}] = i\hbar$ and
$\op{R_A} |r\rangle = r|r\rangle$.

The apparatus is prepared in an initial state
$|\phi(r_0)\rangle$ which is a packet of $|r\rangle$ states,
centered at $r=r_0$. The initial state of the system $|\psi_s\rangle$ can
be expanded in terms of the eigenstates of $\op{Q_S}$, 
$|\psi_s\rangle = \sum_i c_i|s_i\rangle$, where 
$\op{Q_S}|s_i\rangle = s_i|s_i\rangle$.
Let us assume that the measurement interaction is
switched on at $t=0$ and continues till $t=T$, with the proviso
$\int_0^T g(t)dt=1$. The state, after the measurement interaction, is given by
\begin{eqnarray}
|\Psi(T)\rangle &=& e^{-{i\over\hbar} Q_S Q_A}|\psi_s\rangle |\phi(r_0)\rangle
~\nonumber\\
|\Psi(T)\rangle &=& \sum_{i} e^{-{i\over\hbar} s_i Q_A } c_i |s_i\rangle
|\phi(r_0)\rangle \nonumber\\
|\psi(T)\rangle &=& \sum_{i} c_i |s_i\rangle |\phi(r_0+ s_i )\rangle.\nonumber\\
&=& c_1 |s_1\rangle |\phi(r_0+ s_1 )\rangle +
c_2 |s_2\rangle |\phi(r_0+ s_2 )\rangle\nonumber\\
&&+
c_3 |s_3\rangle |\phi(r_0+ s_3 )\rangle \dots
\label{strong}
\end{eqnarray}

The measurement interaction results in various eigenstates of $\op{Q_S}$
becoming entangled with packets of $|r\rangle$ states of the apparatus,
localized at different values of $r$. A narrow packet localized at
$r_0+s_k$, for example, would imply a measured eigenvalue $s_k$ of the system.
However, at this stage there is not one outcome, but a superposition of
various outcomes, with different probabilities.

\subsection{Quantum measurement of single systems}

The preceding discussion of quantum measurement has an interesting consequence
for single systems, i.e., systems for which an ensemble of identical copies
is not available. If the state is a priori unknown, and an ensemble of
identical copies is available, one can perform many measurements on 
different copies, and from the resulting probabilities of various outcomes
$|c_i|^2$, try to infer the values of $C_i$ and reconstruct the original
state using $|\psi_s\rangle = \sum_i c_i|s_i\rangle$.

However, if there is only a single system available, one can choose to
make one measurement, which will give one a single eigenvalue (say) $s_k$,
and the state would have collapsed to $|s_k\rangle$. The measurements 
gives absolutely no information regarding the original state $|\psi_s\rangle$.
These means that if the state of a single system is unknown, it will 
always remain unknown. This is something profound and implies that an
unknown reality cannot be unveiled, even in principle.

The expectation value of any observable always has a well defined value in
any state. But the question that arises is whether expectation value has
any meaning for a single quantum system. Since the only interpretation of
the expectation value traditionally understood is in terms of repeated 
measurements on an ensemble, the answer seems to be that expectation
value has no meaning for a single quantum system.

However, if one could somehow {\em measure} the expectation value of
an observable in a single measurement, one could argue that it has a
meaning. If one could measure the expectation value of an unknown state,
it would imply that the expectation value has an objective reality,
and would probably lend credence to the objective reality of the
quantum state itself.

\section{Protective measurements}

About twenty two years back Aharonov, Anandan and Vaidman (AAV)
proposed a quantum measurement scheme
involving very weak and adiabatic measurements, which they called
``protective'' measurements,
\cite{AV,AAV1,AAV2,metastable,NDHTQ,VAIDMAN,GAO} where they claimed the
possibility of actually measuring the expectation value of any observable in
a restricted class of states.
The proposal initially raised surprise and scepticism among many
\cite{SCHWINGER,UNRUH,ROVELLI,GHOSE,DU,ALTER1,ALTER2,SAM}.  

While conventional quantum measurements are considered strong and 
impulsive, protective measurements make use of the opposite
limit where the coupling between the system and the apparatus is {\it
weak} and {\em adiabatic}. For protective measurements to work, the system
should be in a non-degenerate eigenstate of its Hamiltonian. The interaction
should be so weak and adiabatic that one cannot neglect the free Hamiltonians. Let the Hamiltonian of the combined system be
\begin{equation}
\op{H}(t) = \op{H_A} + \op{H_S} + g(t)\op{Q_A} \op{Q_S}, \label{H_full}
\end{equation}
where various entities have the same meaning as in the preceding section.
The coupling $g(t)$ acts for a long time $T$ and is switched on and
switched off smoothly.  The interaction is 
normalized as $\int_0^T dt g(t) = 1$, and is assumed to be small and
almost constant for the most part, justifying the approximation,
$g(t) \approx 1/T$.
If $|\Psi(0)\rangle$ is the state vector
of the combined apparatus-system just before the measurement process
begins,  the state vector after T is given by
\begin{equation}
|\Psi(T)\rangle = {\cal T} e^{-{i\over\hbar}\int_0^T H(\tau) d\tau} |\Psi(0)\rangle,
\label{psiT}
\end{equation}
where ${\cal T}$ is the time ordering operator. Since the time dependence
of the Hamiltonian is trivial, we may divide the interval
$[0,T]$ into $N$ equal intervals $\Delta T$, so that $\Delta T = T/N$. 
Since the full Hamiltonian commutes with itself at different times during
$[0,T]$, one can write (\ref{psiT}) as
\begin{eqnarray}
|\Psi(T)\rangle = \left[exp[-{i\Delta T\over\hbar}(H_A + H_S +
{1\over T}Q_A Q_S)]\right]^N |\Psi(0)\rangle. \nonumber\\
\end{eqnarray}

In order to solve the dynamics, one has to worry about whether different
operators sitting in the exponential commute with each other or not.
Since designing the apparatus is in the hands of the experimenter, we 
consider the case when $\op{Q_A}$ commutes with the free
Hamiltonian of the apparatus, i.e., $[\op{Q_A},\op{H_A}]=0$, so that we can have
eigenstates $|a_i\rangle$ such that $\op{Q_A} |a_i\rangle = a_i |a_i\rangle$
and $\op{H_A} |a_i\rangle = E_i^a |a_i\rangle$. 
The operators of the system, $Q_S,H_S$, may or may not commute with each other,
the energy eigenstates of the system are given by
\begin{equation}
H_S |\mu\rangle = \mu|\mu\rangle.
\end{equation}
The states $|a_i\rangle$ are also the exact eigenstates of the instantaneous
Hamiltonian $H(t)$, in the apparatus subspace. So, the exact
instantaneous eigenstates can be written in a factorized form
$|a_i\rangle \overline{|\mu\rangle}$ where $\overline{|\mu\rangle}$ are
defined by
\begin{equation}
(H_S + {1\over T}a_i Q_S) \overline{|\mu\rangle}
=\overline{\mu}\overline{|\mu\rangle}
\end{equation}
The system states $\overline{|\mu\rangle}$ depend on the eigenvalue of
$Q_A$.  Let us assume the
initial state to be a direct product of a non-degenerate eigenstate of
$H_S$, $|\nu\rangle$, and $|\phi(r_0)\rangle$:
\begin{equation}
|\Psi(0)\rangle = |\nu\rangle |\phi(r_0)\rangle .
\end{equation}
Introducing complete set of  exact eigenstates in the above equation,
the wave function at a time $T$ can now be written as
\begin{equation}
|\Psi(T)\rangle = \sum_{i,\mu} e^{{i\over\hbar}E(a_i,\mu) N\Delta T}
|a_i\rangle \overline{|\mu\rangle}\overline{\langle\mu|} 
|\nu\rangle \langle a_i| |\phi(r_0)\rangle, \label{psiT1} 
\end{equation}
where the exact instantaneous eigenvalues $E(a_i,\mu)$ can be written as
\begin{equation}
E(a_i,\mu) = E_i^a + {1\over T} \overline{\langle\mu|}Q_S\overline{|\mu\rangle}
a_i + \overline{\langle\mu|}H_S\overline{|\mu\rangle}.
\end{equation}

Till this point we have not made any approximations, except for ignoring
the switching on and switching off times. Now if the measurement interaction
is very weak and highly adiabatic, $1/T$ is very small,
so that 
\begin{equation}
\overline{|\mu\rangle} = |\mu\rangle + {\cal O}(1/T) + ...
\end{equation}
In the large $T$ limit, one can assume the states to be unperturbed,
i.e., $\overline{|\mu\rangle} \approx |\mu\rangle$.
The energy eigenvalues now assume the form
\begin{equation}
E(a_i,\mu) \approx E_i^a + {1\over T} \langle\mu|Q_S|\mu\rangle a_i + 
\langle\mu|H_S|\mu\rangle.
\end{equation}
Assuming the states of to be unperturbed and the energy to be first
order in $1/T$ amounts to doing a first order perturbation theory.
In this approximation $\overline{\langle\mu|}|\nu\rangle \approx 0$,
and the sum over $\mu$ disappears and only the
term where $\mu =\nu$ survives. This allows us to write
the apparatus part of the exponent again in the operator form
\begin{equation}
|\Psi(T)\rangle \approx e^{-{i\over\hbar}H_A T-{i\over\hbar}
Q_A\langle Q_S\rangle _\nu -{i\over\hbar}\langle H_S\rangle _\nu T}
|\nu\rangle |\phi(r_0)\rangle .
\end{equation}

Since $\op{Q_A}$ is an operator conjugate to $\op{R_A}$, it will act
as a generator of translation for $|r\rangle$ states.
The second term in the exponent will shift
the center of the packet $|\phi(r_0)\rangle$ by an amount $
\langle\nu|Q_S|\nu\rangle$:
\begin{equation}
|\psi(T)\rangle = e^{-{i\over\hbar}H_A T-{i\over\hbar}\nu T}
|\nu\rangle |\phi(r_0+\langle Q_S\rangle _\nu)\rangle .\nonumber\\
\end{equation}
Thus, at the end of the measurement interaction, the center of the
apparatus packet $|\phi(r_0)\rangle$ shifts by $\langle\nu |Q_S|\nu\rangle$.
The apparatus thus records, not the eigenvalue of $\op{Q_S}$ as
in (\ref{strong}), but its
expectation value in the initial {\em unknown} state $|\nu\rangle$.
Not only that, within this approximation the system and the apparatus are not
entangled.

\subsection{Some clarifications}

Protective measurements were widely misunderstood and resulted in a
lot of criticism. \cite{SCHWINGER,UNRUH,ROVELLI,GHOSE,DU,ALTER1,ALTER2,SAM,UFFINK}
Here we list some features of protective measurements which should clarify
various issues which were raised. 
\begin{itemize}
\item Protective measurements don't require that the state of the system
be a-priori fully known.\\
Example: atom in a trap where one may not know the exact potential, but
does know that the atom will be in the ground state.\\
One may have made made an energy measurement on a system to know that the
system is in a particular energy eigenstate, but without the knowledge
of the Hamiltonian one cannot know what the eigenstate is, and hence
finding the expectation value of an observable is not possible.

\item The shift in the pointer state is proportional to the expectation value of
the observable being measured.

\item The expectation value is obtained in one single measurement, on a
single system.

\item As shown in the preceding analysis, observable whose expectation value
is measured, need not commute with the Hamiltonian of the system. An 
objection with a contradicting claim was raised by Uffink.\cite{UFFINK}
Shan Gao pointed out the flaw in the argument.\cite{GAOONUFFINK}

\item The system is not entangled with the apparatus after the measurement.

\item The state of the system does not change after the measurement (within
the approximation used).

\item Expectation value of another operator can be measured, after the
measurement of one.
\end{itemize}

\subsection{Generalized protective measurements}

It has also been demonstrated that protective measurements can, in principle,
be performed even in the most general case where  
$[\op{Q_A},\op{H_A}]\neq 0$ and $[\op{Q_S},\op{H_S}]\neq 0$. \cite{NDHTQ}
However in actual practice, finding the right observables for the 
apparatus, and satisfying all the constraints may be a formidable challenge.
This is so because in this case, the initial apparatus is not supposed to
be a packet of eigenstates of the operator conjugate to $\op{Q_A}$.
Rather it is supposed to be a packet of eigenstates of the operator
conjugate to an operator $\op{Y}$ define as
\begin{equation}
\op{Y}=\sum_j\langle \op{Q_A}\rangle_{a_j}|{a_j}\rangle\langle {a_j}|,
\end{equation}
where various entities have the same meaning as in the preceding analysis.
Whether such an operator can always be found in practice, is an open
question.

In all this analysis we have not considered the dynamical effect of
the ``free" Hamiltonian of the apparatus. This Hamiltonian will cause
the spreading of the packet of the initial apparatus state. In normal
course of action, one would have ignored this effect. However, since
the protective measurements are supposed to be adiabatic, the effect
of the apparatus Hamiltonian, though small, will be cumulative. In other
words, the pointer packet may spread considerably during the course of
protective measurement interaction. Finding the centre of a large
packet, to read out the measured expectation value, would be a difficult
task, and one may have to apply some special techniques to 
do that. \cite{NDHTQ}

\subsection{Does it really work for a single system?}

The success of protective measurements crucially depends on the assumption
that the entanglement between the system and the apparatus can be made
negligibly small, but the system will still shift the pointer state
by a finite amount.

Let us now quantify the effect of the terms we have neglected till now.
If we consider the state of the system and apparatus to be perturbed
to first order in $1/T$, as opposed to being unperturbed in the preceding
analysis, the final state, to next higher order in $1/T$, would look like
\begin{eqnarray}
|\Psi(T)\rangle &\approx& e^{-{i\over\hbar}\op{H_A} T-{i\over\hbar}\nu T}
{|\nu\rangle} {|\phi(r_0+{\langle Q_S\rangle_\nu})\rangle} \nonumber\\
&&+ {1\over T}\sum_{\mu (\ne\nu)} \alpha_{\mu\nu} e^{-{i\over\hbar}{H_A} T-{i\over\hbar}\mu T}
{|\mu\rangle} {|\phi(r_0+{\langle Q_S\rangle_\mu})\rangle}\nonumber\\
\end{eqnarray}
where
$\alpha_{\mu\nu}$ involve matrix elements of $\op{Q_S}\op{Q_A}$ among
various unperturbed states, and unperturbed energies. The above is an entangled state, and in a real
measurement there is a probability that the original state of the system,
which was $|\nu\rangle$ to begin with, gets changed to (say) $|\mu\rangle$, and
the apparatus state gets shifted by $\langle Q_S\rangle_\mu$. The probability
of this happening is proportional to $1/T^2$. 

One can see that by increasing $T$ and weakening the interaction, the
probability of the protective measurement failing can be made smaller,
but can never be made zero. In general, it has been rigorously shown that the
{\em state disturbance} in protective measurements scales as 
$1/T^2$.\cite{SCHLOSSHAUER} This indicates that although a practical
implementation of protective measurement is possible, for a single unknown
state, one can never be sure that the protective measurement has yielded
the expectation value in the original state. Hence it cannot be used
to argue for a strict objective reality of the wave-function.

\section{Experimental realization}

The idea of protective measurements was proposed more than two decades back,
but an experimental demonstration of the same is still lacking.
The reason for this is that there are several constraints which the system
and the apparatus should satisfy before one can carry out a successful
protective measurement. The adiabatic nature of interaction may also
present some difficulty.

\subsection{Cold atoms for testing protective measurements}

A proposal was made for testing protective measurements using cold atoms
in a Stern-Gerlach like setup.\cite{NDHCOLD} We briefly describe it in the
following. Low velocity of cold atoms
may be exploited for achieving adiabaticity to some degree.
\begin{figure}
\centerline{\resizebox{8.0cm}{!}{\includegraphics{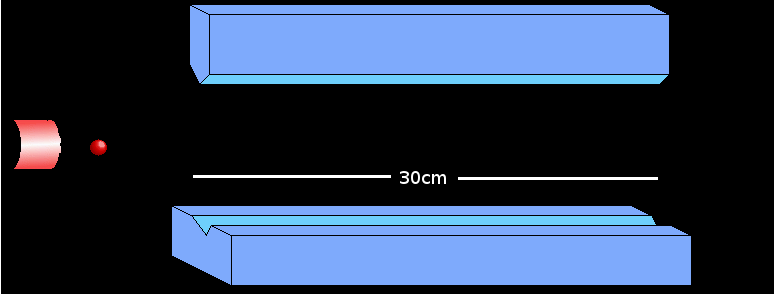}}}
\caption{A schematic diagram of a Stern-Gerlach type experiment with cold atoms.}
\label{cold}
\end{figure}
The Hamiltonian of the atom with mass $m$ and magnetic moment $\mu$, can be
written as
\begin{equation}
\op{H} = {\op{P}^2\over 2M} - \mu B_0\op{\vec{\sigma}}\cdot \vec{n}_0
- \mu g(t)B_i \op{x}\op{\vec{\sigma}}\cdot \vec{n}
\end{equation}

Initial system-apparatus state:
\begin{equation}
|\Psi(0)\rangle = {|+\rangle} {|\phi_p(0,\epsilon)\rangle}
~~~~~~~~~ \vec{\sigma}\cdot \vec{n}_0{|\pm\rangle} = \pm{|\pm\rangle},
\end{equation}
where $|\phi_p(0,\epsilon)\rangle$ is a Gaussian wave-packet in the momentum
space, with zero average momentum, and a width $\epsilon$ in momentum space.
The state after a time $T$ is given by
\begin{equation}
|\Psi(T)\rangle = e^{-iHT}{|+\rangle} {|\phi_p(0,\epsilon)\rangle} 
\end{equation}
The position operators $\op{x}$ will act as a generator of translation 
in momentum space, and the system-apparatus state, at the end of the
measurement interaction, is given by
\begin{eqnarray}
|\Psi(T)\rangle &\approx& e^{-iP_x^2T/2M}e^{-i\mu B_0T} e^{\mu B_i\vec{n_0}\cdot\vec{n}x}{|+\rangle} {|\phi_p(0,\epsilon)\rangle} \nonumber\\
&=& e^{-i\mu B_0T} {|+\rangle} {|\phi_p({\langle\mu B_i\vec{\sigma}\cdot\vec{n}\rangle_{+}},\epsilon(T))\rangle} 
\end{eqnarray}
The state of the spin and the state of the atom are disentangled, and
${|\phi_p({\langle\mu B_i\vec{\sigma}\cdot\vec{n}\rangle_{+}},\epsilon(T))\rangle}$ 
is a Gaussian with a momentum ${\langle\mu B_i\vec{\sigma}\cdot\vec{n}\rangle_{+}}$,
with a width $\sqrt{\epsilon^2 + {T^2\over M^2\epsilon^2}}$

If one uses the following experimental parameters:
$\epsilon = 1$ mm, $L = 30$ cm, $B_0 = 1$ Gauss, atom velocity:
$v \sim 1$ cm/s,

Momentum shift $\gg$ Momentum spread

As the atom travels after coming out of the interaction region, it position
will shift, because of the non-zero average momentum.
Position shift after 30 sec evolution after interaction will about about 2 cm.

\section{Conclusions}

We have described protective measurements which are a promising tool for
probing single systems, i.e., systems for which an ensemble of identical
quantum states is not available. Protective measurements lend a new
experimental meaning to the quantum expectation value which, traditionally,
has meaning only in the context of many eigenvalue measurements over an
ensemble. Although protective measurements can be practically used to
measure the expectation value in a single measurement, the non-zero error
which is always present, rules out using the same to assign any object
reality to the wave-function. An experimental test of protective measurements
should be possible, and one proposal for the same has been described here.

\end{document}